\documentclass{svproc}
\usepackage{braket}
\usepackage{xcolor}
\usepackage{bm}
\usepackage{amsmath}
\usepackage{graphicx}
\usepackage{multicol}
\usepackage{footmisc}
\usepackage{url}

\begin{document}
\mainmatter              
\title{Linking nuclear reactions and nuclear structure to study exotic nuclei using the dispersive optical model}
\titlerunning{Dispersive Optical Model}  
%
\author{W. H. Dickhoff}
\authorrunning{Dickhoff} 
%
%
\institute{Department of Physics, Washington University, St. Louis MO 63130, USA,\\
\email{wimd@physics.wustl.edu},\\ WWW home page:
\texttt{https://web.physics.wustl.edu/wimd/}}

\maketitle              

\begin{abstract}
The dispersive optical model (DOM)  is employed to simultaneously describe elastic nucleon scattering data for ${}^{40}$Ca, ${}^{48}$Ca, and ${}^{208}$Pb as well as observables related to the ground state of these nuclei, with emphasis on the charge density.
Such an analysis requires a fully non-local implementation of the DOM including its imaginary component.
Illustrations are provided how ingredients thus generated provide critical components for the description of the $(d,p)$ and $(e,e'p)$ reaction.
For the nuclei with $N > Z$ the neutron distribution is constrained by available elastic scattering and ground-state data thereby generating a prediction for the neutron skin.

We identify ongoing developments including a non-local DOM analysis for ${}^{208}$Pb and point out 
possible extensions of the method to secure a successful extension of the DOM to rare isotopes. 
\end{abstract}
\section{Introduction}

How do the properties of protons and neutrons in the nucleus change from the 
valley of stability to the respective drip lines?
The answer can be developed by studying  the propagation of a nucleon through the nucleus at positive energy, generating experimentally accessible elastic scattering cross sections, as well as the motion of nucleons in the ground state at negative energy.
The latter information sheds light on the density distribution of both protons \emph{and} neutrons relevant for clarifying properties of neutron stars. 
Detailed knowledge of this propagation process allows for an improved 
description of other hadronic reactions, including those that purport to 
extract structure information, like transfer or knockout reactions.
Structure information associated with the removal of nucleons
from the target nucleus, is therefore subject of these 
studies and must be supplemented by the appropriate description of the 
hadronic reaction utilized to extract it.
Consequently, a much tighter link between reaction and structure studies than is common practice is an 
important goal of this research. 

In our group we apply the Green's functions method~\cite{Dickhoff04,Dickhoff08} to the nuclear many-body problem to address this issue with special emphasis on reaching the limits of stability. 
The method can be utilized to correlate huge amounts of experimental
data, like elastic nucleon cross sections, analyzing powers, etc.,
as well as structure information like removal energies, density 
distributions, and other spectral properties.
This is achieved by relating these data to the nucleon self-energy employing its causal properties in the form of a subtracted dispersion relation.
The current implementation and corresponding details can be found in~\cite{Dickhoff17}. 
The method is known as the dispersive optical model (DOM) and has proceeded way beyond its original form~\cite{Mahaux:1991}.
A more general review of the optical model is available in~\cite{Dickhoff:2019}.
We discuss some recent developments of the DOM with applications to transfer reactions in Sec.~\ref{sec:tr}, the analysis of the $(e,e'p)$ reaction solely with DOM ingredients in Sec.~\ref{sec:eep}, predictions of neutron distributions in Sec.~\ref{sec:skin}, and finally offer some conclusions in Sec.~\ref{sec:con}.

\section{Transfer reactions and the DOM}
\label{sec:tr}
Transfer reactions are under intense study in order to develop a reliable method to generate accurate results given certain ingredients like overlap functions and optical potentials.
A remaining source of uncertainty in the calculation of transfer
reaction observables is the optical potential for the relevant nucleons and the deuteron.
Our group has made several contributions to this effort documented in Refs.~\cite{Nguyen11,Ross15} mostly involving exploratory efforts.

Deuteron-induced reactions have played an important role in elucidating properties of neutrons that are either added to or removed from the target nucleus. This role will be even more prominent when such transfer reactions are studied in inverse kinematics at radioactive beam facilities like FRIB~\cite{Bardayan:2016,Wimmer:2018}. While scientifically compelling in its own right, the $(d,p)$ reaction also yields indirect access~\cite{Escher12} to the study of neutron capture and therefore provides essential information for the $(n,\gamma)$ reaction which is critical for the study of the understanding of the $r$-process.

The present state of the reaction description can be summarized by noting that the distorted-wave Born approximation and coupled-channel approaches have mostly studied discrete final states. The treatment of the continuum was proposed in the late seventies but efforts ended in the nineties, with an unresolved controversy. Only recently, three different groups~\cite{Potel15,Lei15,Carlson15} have revived this subject and during a recent workshop at MSU/FRIB~\cite{dp16} have concluded that the relevant issues have now been resolved.

\begin{figure}[tpb]
\begin{center}
\includegraphics[scale=.35]{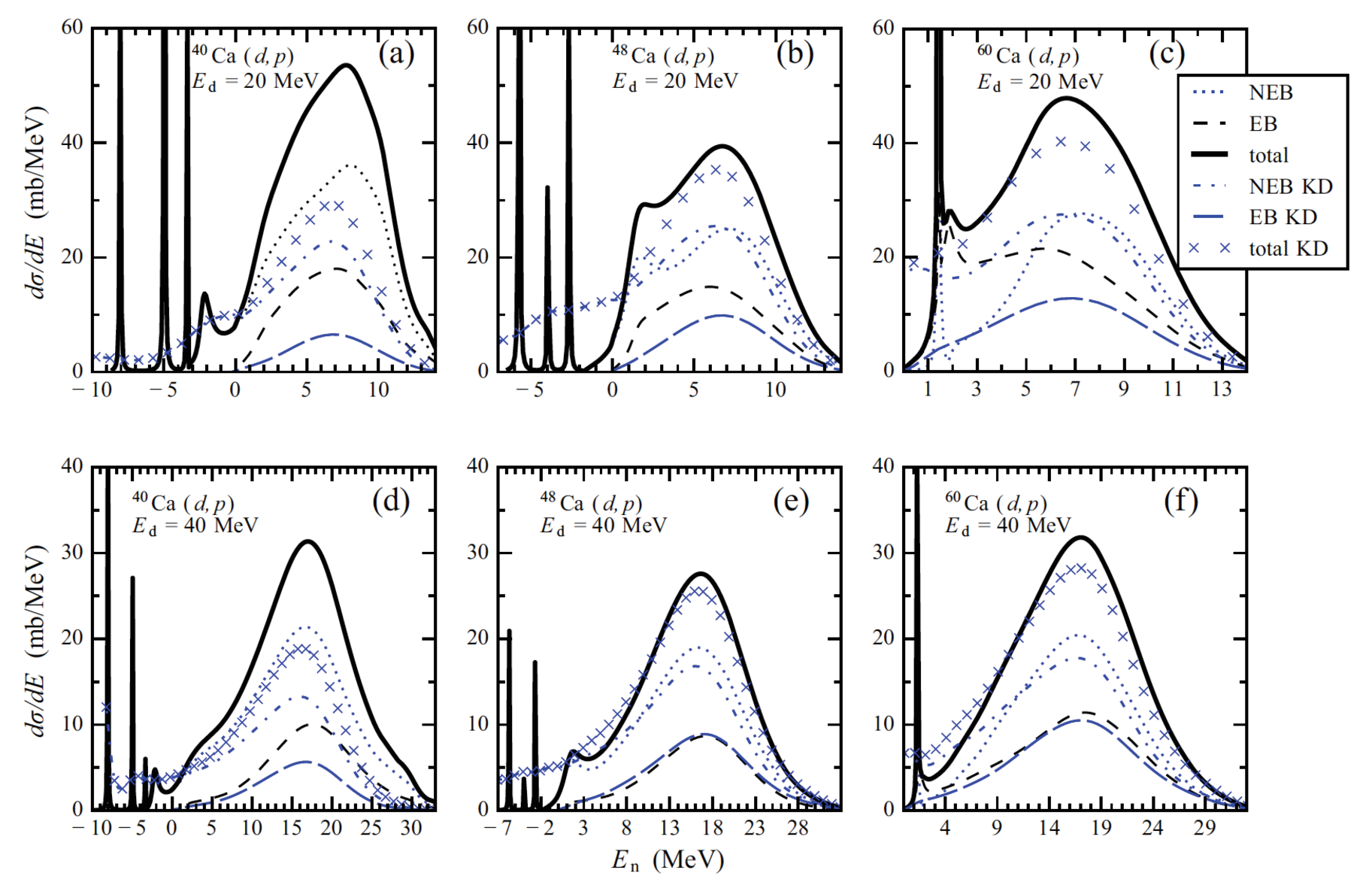}
\caption{Comparison of KD phenomenological optical potential and the DOM~\cite{Potel17}: elastic breakup (EB) and non-elastic breakup (NEB) proton spectra for the reactions ${}^{40}$Ca$(d, p)$, ${}^{48}$Ca$(d, p)$, and ${}^{60}$Ca$(d, p)$ at $E_d$ = 20 MeV and $E_d$ = 40 MeV.}
\label{fig:dp}
\end{center}
\end{figure}

The main ingredients of the present state of the $(d,p)$ reaction description allows a simultaneous treatment of transfer, elastic breakup, and the formation of the compound nucleus. Critical ingredients for the relevant calculations are provided by the deuteron optical potential, the description of the propagation of the added neutron, and the final proton optical potential.  
Phenomenological optical potentials suffer from being non-dispersive, local, and are not constrained by negative energy data. 
A proper description of the reaction therefore requires dispersive, non-local potentials that are also constrained by negative energy data. 
Such potentials are provided by the latest implementation of the DOM~\cite{Dickhoff17} for the neutron and proton propagation.
An initial assessment of the DOM ingredients has been implemented by employing the local version~\cite{Charity11} for Ca isotopes including an extrapolation to ${}^{60}$Ca. 
These results together with an overview of the current theory relevant for elastic and non-elastic breakup have been published in~\cite{Potel17}.
Already at this early stage, a clear preference of DOM-generated potentials emerges over a more traditional global optical potential like the one of~\cite{Koning03} labeled KD, 
as illustrated in Fig.~\ref{fig:dp}.

As the DOM potentials are constructed to smoothly connect the positive and negative energy domain, they accurately describe the peaks that occur when a neutron is added in a bound state, whereas phenomenological potentials do not provide a suitable extrapolation to negative energy.
Available data are well described with these potentials~\cite{Potel17}. 
Further developments are necessary to raise the standard for the description of the deuteron and employ non-local dispersive potentials for nucleons in order to analyze data from this reaction employing rare isotopes in inverse kinematics.
The main missing ingredient is an appropriate description of the deuteron for which only local, non-dispersive potentials are available~\cite{An:2006,Han:2006,Zhang:2016}.
We are presently developing tools to describe the deuteron by a non-local, dispersive potential that is constrained by corresponding elastic scattering data. 
The proposed approach depends on recognizing that elastic deuteron scattering can be interpreted as the propagation of an interacting proton-neutron pair in the medium provided by the target nucleus~\cite{vill67}.

\section{${}^{40}$Ca$(e,e'p)^{39}$K reaction and spectroscopic factors}
\label{sec:eep}

Several papers have appeared in the past questioning the relevance of spectroscopic factors~\cite{Zhanov10,Furnstahl10} and the possibility of measuring momentum distributions or occupation numbers~\cite{Furnstahl02}.
It is useful to point out that Fermi liquid theory developed by Landau~\cite{Lan57a,Lan57b,Lan59} relies on the notion of a quasiparticle with a corresponding strength (spectroscopic factor) near the Fermi surface that can be experimentally probed through specific heat measurements~\cite{Wheatley75}.
For finite systems like atoms and molecules the corresponding information is accessed by analyzing the $(e,2e)$ reaction~\cite{Dickhoff08,Lohmann81,McCarthy91}.
Similar efforts in nuclear physics have attempted to extract spectroscopic factors from the $(e,e'p)$ reaction~\cite{Lapikas93} for valence hole states in mostly double-closed-shell nuclei (see also Refs.~\cite{Dickhoff08,Dickhoff10}).

Experimental results of the $(e,e'p)$ reaction have been included in the local DOM in the past by employing the extracted spectroscopic factors~\cite{Kramer89,Kramer01} in fits with local potentials to the ${}^{40}$Ca and ${}^{48}$Ca nuclei~\cite{Charity06,Charity07} and to data in other domains of the chart of nuclides~\cite{Charity11}.
A better approach has now been implemented based on the non-local DOM developments~\cite{Dickhoff17,Mahzoon14,Mahzoon:2017} that also allows an assessment of the quality of the distorted-wave impulse approximation (DWIA) that is utilized to describe the reaction.
We note that the conventional analysis of the reaction employed standard local non-dispersive optical potentials to describe the proton distorted waves~\cite{denHerder88}.
We have thus arrived at a stage with the DOM that all ingredients for the DWIA description can be supplied from one self-energy that generates the proton distorted waves at the desired outgoing energies, as well as the overlap function with its normalization.
Important to note is that these ingredients are not adjusted in any way to $(e,e'p)$ data.

The non-local DOM description of ${}^{40}$Ca data was presented in~\cite{Mahzoon14}.
In the mean time, additional experimental higher-energy proton reaction cross sections~\cite{PhysRevC.71.064606} have been incorporated which caused some adjustments of the DOM parameters compared to~\cite{Mahzoon14}. 
Adjusting the parameters from the previous values~\cite{Mahzoon14} to describe these additional experimental results leads to an equivalent description for all data except these reaction cross sections.
The required additional absorption at higher energies leads to a loss of strength below the Fermi energy, reducing the spectroscopic factors by about 0.05 compared to the results reported in~\cite{Mahzoon14}, thereby also documenting the importance of reaction cross section data for protons at higher energy.

\begin{figure}[tpb]
\begin{center}
\includegraphics[scale=.4]{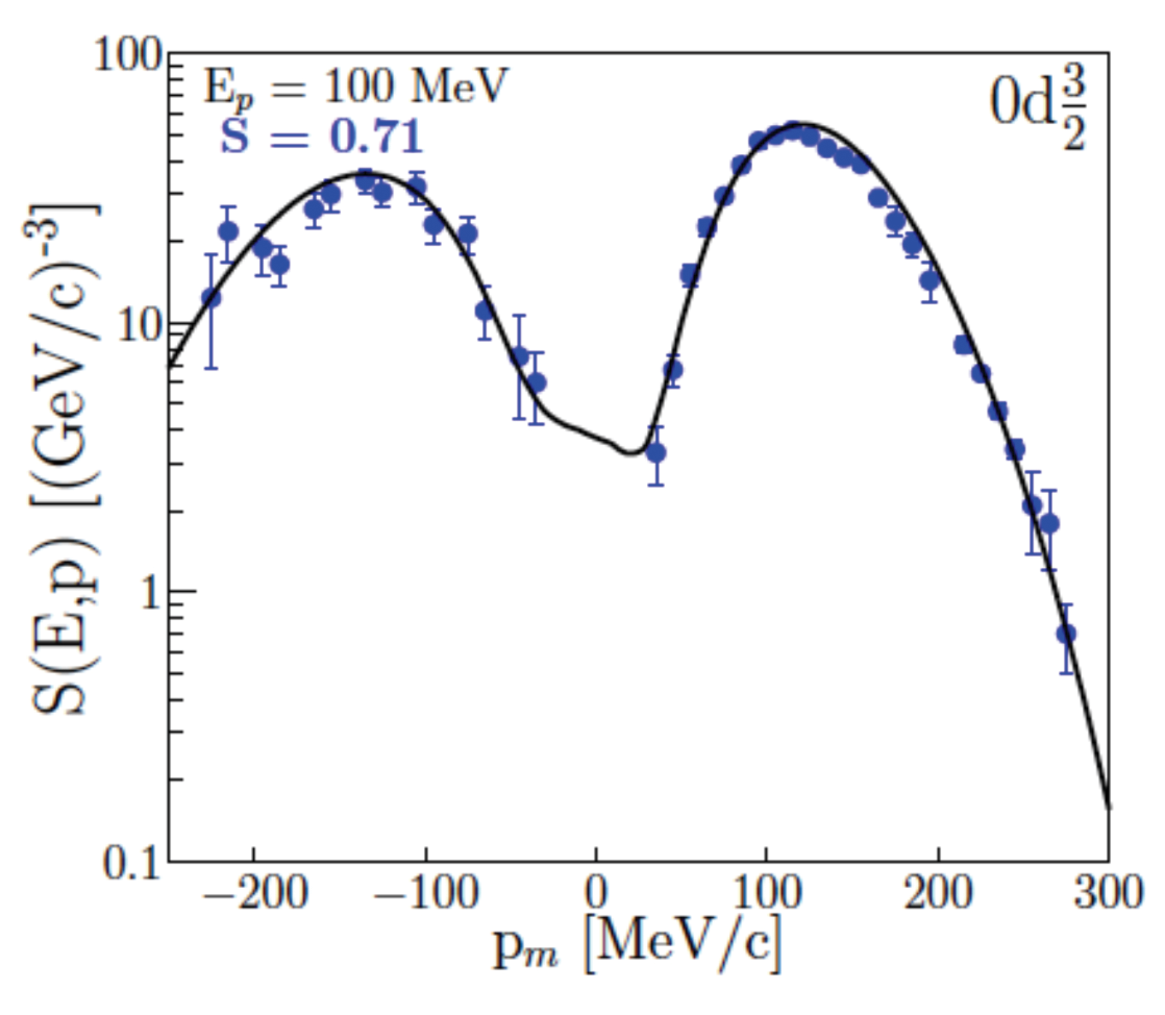}
\caption{Comparison of the spectral distribution measured at Nikhef for outgoing proton energies of 100~MeV to DWIA calculations using the proton distorted waves, overlap function and its normalization from a non-local DOM parametrization.
Results are shown for the knockout of a 0$\textrm{d}_{3/2}$ proton from ${}^{40}$Ca to the ground state of ${}^{39}$K.
} 
\label{fig:mack5}
\end{center}
\end{figure}

Using a recent version of the code DWEEPY~\cite{Giusti11}, 
our DOM ingredients have been utilized to describe the knockout of a proton from the $0\textrm{d}\frac{3}{2}$ and $1\textrm{s}\frac{1}{2}$ orbitals in ${}^{40}$Ca with fixed normalizations of 0.71 and 0.60, respectively~\cite{Atkinson:2018}. 
The DOM at present generates only one main peak for $1\textrm{s}\frac{1}{2}$ orbit so the employed value of 0.60 for the spectroscopic factor takes into account the experimentally observed low-energy fragmentation.
Experimental data were obtained at Nikhef in parallel kinematics for three outgoing proton energies: 100, 70 and 135 MeV.
Data for the latter two energies were never published before.
The resulting description of the $(e,e'p)$ cross sections is at least as good as the Nikhef analysis which yielded spectroscopic factors of 0.65$\pm$0.06 and 0.51$\pm$0.05 for these orbits at 100 MeV~\cite{Kramer89}, as illustrated in Fig.~\ref{fig:mack5}.
Our results demonstrate that the DWIA reaction model is still satisfactory at 70 MeV and 135 MeV outgoing proton energies.
By applying the bootstrap method used for the neutron skin calculation of~\cite{Mahzoon:2017}, we have generated errors for the spectroscopic factors for these orbits with values 0.71$\pm$0.04 and 0.60$\pm$0.03, for the $0\textrm{d}\frac{3}{2}$ and $1\textrm{s}\frac{1}{2}$ orbitals in ${}^{40}$Ca, respectively.
The results further suggest that the chosen window around 100 MeV proton energy provides the best and cleanest method to employ the DWIA for the analysis of this reaction. 

We therefore make a strong case that the canonical suppression of the spectroscopic factors as pioneered by the Nikhef group~\cite{Lapikas93} continues to generate values of around 0.7 although there are qualitative differences in the construction of the cross sections on account of the non-local potentials that determine the distorted proton waves.
Further insight into the claim that the $(e,e'p)$ reaction can yield absolute
spectroscopic factors for low-lying discrete states in the final 
nucleus~\cite{Sick91,Pandharipande97,Dickhoff10} has therefore been provided, while demonstrating that a consistent description of the reaction ingredients as provided by the non-local DOM is essential.

\section{Neutron distributions and the DOM}
\label{sec:skin}

The efficacy of the DOM has recently been documented when its fully non-local implementation was extended to ${}^{48}$Ca.
Available ground-state properties of ${}^{48}$Ca appropriate for a study of the properties in this system, apart from the important particle numbers of $Z = 20$ and $N = 28$, include the charge density in addition to level structure. 
These properties on top of the standard elastic scattering data available at positive energy have been employed to construct the $N-Z$ dependence of the DOM potential leaving all ingredients of the fit to ${}^{40}$Ca fixed except for radius parameters.
Excellent agreement with the experimental charge density has been obtained~\cite{Mahzoon:2017} just as earlier for ${}^{40}$Ca~\cite{Mahzoon14}. 

Recently acquired elastic neutron scattering data and total cross sections for ${}^{48}$Ca were published earlier in our large DOM paper~\cite{Charity11} but it was at that time not possible to generate an accurate fit to the differential cross sections at low energy employing the local implementation of the DOM.
Our current non-local DOM potentials provide increased flexibility that allows for the present excellent fit to these data. 
Most of the properties of the first 20 neutrons in this nucleus are already well-constrained by the fit to the properties of ${}^{40}$Ca.
The additional influence of the extra 8 neutrons in this nucleus is then further constrained by these elastic scattering data and total neutron cross sections~\cite{Charity11} as well as level structure.
The neutron properties of ${}^{48}$Ca are of extreme interest to the community since the neutron radius can be experimentally probed without ambiguity employing parity-violating elastic electron scattering experiments at Jefferson Lab~\cite{CREX13}.

\begin{figure}[tpb]
\begin{center}
\includegraphics[scale=.6]{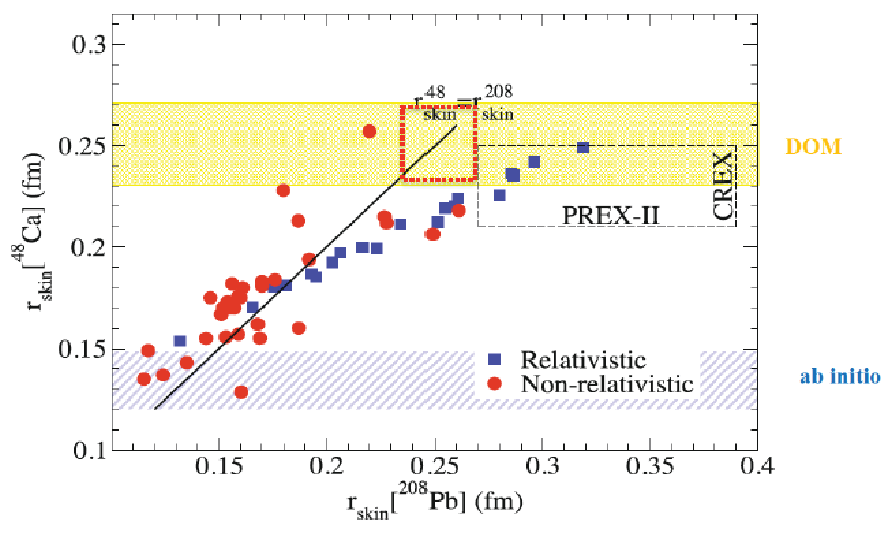}
\caption{Figure adapted from~\cite{Horowitz14} with the results from Refs.~\cite{Hagen16} and \cite{Mahzoon:2017} indicated by horizontal bars relevant for ${}^{48}$Ca and the big dotted square including the preliminary DOM result for ${}^{208}$Pb. Smaller squares and circles refer to relativistic and nonrelativistic mean-field calculations cited in Ref.~\cite{Horowitz14}.}
\label{fig:skin}
\end{center}
\end{figure}
To produce a theoretical error for our result for the neutron skin 
we have employed a method that was explored in the determination of the Chapel-Hill global optical potential~\cite{Varner91}.
These results have now been published in~\cite{Mahzoon:2017} with our neutron skin prediction of 0.249$\pm$0.023 fm which is much larger than the prediction of the \textit{ab initio} coupled-cluster calculation reported in~\cite{Hagen16} and most mean-field calculations~\cite{Horowitz14}.
We note that this work fulfills the earlier promise of the DOM, in that it can be employed to make sensible predictions of important quantities constrained by other experimental data.
When envisaged earlier~\cite{Charity06}, it was thought that these predictions would involve only rare isotopes but important quantities for stable nuclei also fall under its scope.  
We show in Fig.~\ref{fig:skin} results for the neutron skin of ${}^{48}$Ca plotted versus the one of ${}^{208}$Pb as presented in~\cite{Horowitz14}, while adding horizontal bars for the DOM result~\cite{Mahzoon:2017} and the coupled-cluster result of~\cite{Hagen16}.
Our current efforts for ${}^{208}$Pb are also generating a large neutron skin as indicated by the large square in Fig.~\ref{fig:skin}.
The dashed box includes the central value of~\cite{PhysRevLett.108.112502} but with the expected error of the PREX-II experiment. 
The expected error for the CREX experiment~\cite{CREX13} is indicated by the vertical width of the box while its central value is arbitrarily chosen.

\section{Conclusions}
\label{sec:con}
As illustrated in this paper, the DOM provides ingredients for transfer reactions, the $(e,e'p)$ reaction, and predictions for the neutron skin of ${}^{48}$Ca and ${}^{208}$Pb, demonstrating the relevance of this approach to simultaneously answer the questions how nucleons propagate through the nucleus at positive energy and where they are localized in the ground state. 
Extensions to other knockout reactions like $(p,pN)$ and the improved description of the deuteron will likely contribute to a robust extension of the DOM to rare isotopes.

\section*{Acknowledgements}
   This work was supported by the U.S. National Science Foundation under grant PHY-1613362 and contains critical contributions of Hossein Mahzoon and Mack Atkinson as part of their thesis research. Contributions of other collaborators are also gratefully acknowledged.
%
%
\bibliographystyle{spphys}
\bibliography{opt}
\end{document}